\documentclass[pre,reprint, showpacs]{revtex4-1}

\usepackage{graphicx}
\usepackage{amssymb}
\usepackage{amsmath}
\usepackage{longtable}
\graphicspath{{./Figures.eps/}}

\begin{document}
 \author{H. Souguir, O. Ronsin, C. Caroli, and T. Baumberger}
 \affiliation{Institut des nanosciences de Paris (INSP), Universit\'e Paris 6, UMR CNRS 7588,
2 place Jussieu, 75005 Paris, France}
\email{tristan@insp.jussieu.fr}

\title{Two-step build-up of a thermoreversible polymer network: \\From early local to late collective dynamics}

\begin{abstract}

{We probe the mechanisms at work in the build-up of   thermoreversible   gel networks,  with the help of hybrid gelatin gels containing a controlled density of irreversible, covalent crosslinks (CL), which we quench below the physical gelation temperature. The detailed analysis of the dependence on covalent crosslink density of both the shear modulus and optical activity evolutions with time after quench enables us to identify two stages of the physical gelation process, separated by a temperature dependent crossover modulus:
(i)  an early nucleation regime during which rearrangements of the triple-helix CL play a negligible role,  (ii)  a late, logarithmic aging one, which is preserved, though slowed down, in the presence of irreversible CL.  We show that aging is fully controlled by rearrangements and discuss the implication of our results in terms of the switch from an  early, local dynamics to a late, cooperative long-range one. } 

 \end{abstract}
 

\maketitle

\section{Introduction}
Disordered solids, namely various glasses, polymeric or not, exhibit, following fast relaxation after quench, a slow, usually quasi-logarithmic, mechanical strengthening. This self-decelarating behavior is known as aging. As pointed out by Parker and Normand\cite{Parker}, this specificity of glassy dynamics, commonly assigned to the complexity of the free energy landscape, is shared by another class of materials, namely thermoreversible hydrogels. Indeed, although their polymer content is in general very small (typically a few percent), when they are quenched below the gelation temperature $T_g$, as crosslinks proliferate beyond percolation, the corresponding topological constraints increasingly decimate the paths of easy relaxation. 

While logarithmic aging has been evidenced in numerous gels \cite{Nij}, extensive systematic studies have focussed on gelatin ones. Gelatin consists of polypeptide chains which self-assemble, when quenched below $T_g\simeq 30^\circ$C, via partial renaturation of the native collagen triple-helix  structure \cite{Ferry, Harrington} from the high temperature free chain coil   configuration. The chirality of the resulting crosslinks (CL) permits to characterize this gelation dynamics by monitoring the evolution, not only of the storage shear modulus $G$, but also the optical activity, hence the renatured collagen fraction $\chi$ \cite{Papon}. 

There is general agreement \cite{Nij, Guo} on the fact  that gelation proceeds via the following scenario:  

(i) Formation of triple-helix nuclei whose critical size decreases as temperature  $T$ decreases below $T_g$; 

(ii) subsequent extension (growth) of these critical crosslinks while nucleation gradually slows down as the network mesh size decreases. 

\begin{figure}[htbp]
\begin{center}
\includegraphics[width = 8 cm]{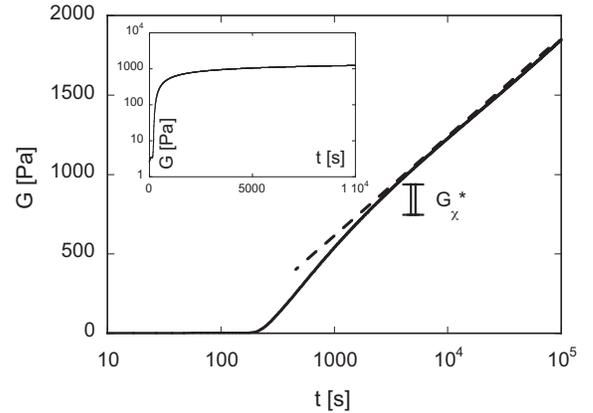}
\caption{Evolution of the shear modulus of a 5 wt.\% gelatin gel in water at $T = 20^\circ$C, vs. logarithm of time after quanch.  The dashed line fits the late  aging behavior. The double bar indicates the crossover region as deduces from optical data (see text section III.A.). Insert: same data in the lin-log representation. }
\label{Fig:loglin}
\end{center}
\end{figure}

\noindent
The associated depletion of the entropy supply of the polymer chains leads to gradual exhaustion of these nucleation and growth processes, hence to a severe slowing down of the growth of $G$. This is often described as ``saturation of the modulus'', since many authors  traditionally resort to the $\log G$ vs. $t$ plot an example of which is shown on Fig.\ref{Fig:loglin} (insert). 
However, in this late regime, as illustrated on Fig.\ref{Fig:loglin}:

(iii) structural relaxation does not stop but crosses over to the self-decelerating, logarithmic dynamics ($G\sim \log t$) characteristic of glass aging. No trend toward saturation has been observed up to two months \cite{NormandAging}. 

This aging behavior,  specific of physical  (as opposed to covalent) gels, has been assigned to network rearrangements taking place via thermally activated zippings and unzippings of the H-bonds stabilized triple helices, i.e. to the thermoreversibility of the CL. However, no direct information about the relative weight of nucleation/growth versus ``crosslink plasticity'' in the early and late regimes of network build-up has been available up to now. 
 In order to try and shed more light on this issue, we resort in this work to an approach which parallels the study of  point-to-set correlations in glass-forming liquids \cite{Kob}.
 That is, in order to assess the influence of reversibility on the dynamics, we introduce in the gels a controlled density of covalent CL. 
 
 More precisely, we first form at a temperature above $T_g$  a covalent network, characterized by its shear modulus $G_{ch}$. Further covalent crosslinking is then inhibited, after which we quench the gel below $T_g$ so as to allow for physical crosslinking to take place. This leads to the formation of  ``chemical-physical hybrid gels''  \cite{Larreta, Ross} the aging dynamics of which we monitor via the evolution of their modulus $G(t)$ and  renatured collagen fraction $\chi(t)$. 

 As  can be intuitively expected, increasing $G_{ch}$, hence the fraction of irreversible CL, results in a slower and slower dynamics. Moreover, the qualitative features of  $\chi(t)$ and $G(t)$ --- including non-saturating, logarithmic aging of the shear modulus ---  are fully preserved. Detailed quantitative analysis of the dependence upon $G_{ch}$ of network build-up  dynamics enables us to identify  a temperature-dependent cross-over shear modulus  separating (i) an early nucleation regime during which crosslink rearrangements play a negligible role,  from (ii)  the late, logarithmic aging one, fully controlled by rearrangements. We discuss the implication of our results in terms of the switch from an  early, local dynamics to a late, cooperative long-range one.  

\section{Materials and methods}
\paragraph*{Sample preparation --- }
The first stage   consists in the preparation of a covalently bonded gelatin network with a given modulus $G_{ch}$. For this purpose we use an enzymatic  route which we have  described in detail  elsewhere \cite{GelZ}.

In short, we dissolve gelatin (300 Bloom, type A from porcine skin, Sigma) in deionized water et 65$^\circ$C. After total dissolution of the polymer,  the solution is quickly mixed  at 40$^\circ$C with a Tgase enzyme solution (microbial transglutaminase, Activa-WM, Ajinomoto Foods Europe SAS) so as to reach a final composition of 5 wt\% gelatin and 2.6 nmol of Tgase (corresponding to an enzymatic activity of 2U). The solution is poured into the measurement cells of  both  a rheometer and a polarimeter (see below).  Gelation then proceeds at $T_{set} = 40^\circ$C. At this temperature, chosen well above $T_g \simeq 30^\circ$C, no triple-helix {\it reversible} cross-link can form whereas Tgase catalyzes actively  the formation of inter-chain {\it covalent} bonds between two specific residues \cite{Tgase}. 

When the target shear modulus level, as monitored in the rheometer, is reached, the enzyme is inhibited by quickly heating both samples up to 70$^\circ$C and staying at this temperature for 10 min, after which they are cooled down back to 40$^\circ$C. We have checked that the shear modulus, measured at this temperature, remains constant over days.

The physical gelation is subsequently triggered by quenching to the working temperature $T<T_g$ --- in most cases, $T = 20^\circ$C.

\paragraph*{Shear modulus measurements ---}

The elastic response of the hybrid gel is characterized in a stress-controlled rheometer (AMCR 501, Anton Paar) equiped with a plane-plane, sand-blasted cell, oscillating at 1 Hz, with a 1\% strain amplitude so as to  operate in the linear response regime. 
 The sample is protected against solvent evaporation  by a paraffin oil rim. We quench the samples at the maximum rate (15$^\circ$C/min) permitted by the  thermoelectric cooling device so as to minimize the formation of physical crosslinks prior to reaching the target temperature $T$.  
 
 We will need to characterize the hybrid gels   by the modulus  of the purely covalent network at  temperature $T$. Since  physical gelation starts without any time-lag (see below), it is not possible to measure reliably $G_{ch}(T)$. Since the gel elasticity is of entropic origin, the shear modulus at fixed network structure scales as the absolute temperature $T$.   So, as a proxy for $G_{ch}(T)$, we use the following expression: 
 $G_{ch}(T) = (T/T_{set})G_{ch}(T_{set})$ with both temperatures in $^\circ$K.

All the  gels studied here exhibit a very small level of viscoelastic losses (typically the ratio of the loss to storage moduli, $G''/G'\lesssim 10^{-2}$ at 1 Hz).  So, for the sake of simplicity we  characterize them by their sole  storage shear modulus, noted $G$.  
\paragraph*{Optical activity measurements ---}

The rotation angle $\alpha$ of the polarization plane of light passing through a 1 cm gelatin sample  is measured using  a spectropolarimeter (Jasco 1100) operating at 436 nm.  Gelatin is optically active in both the random coil (above $T_g$) and helix (below $T_g$) conformations. The renatured collagen  fraction $\chi$ is obtained as \cite{Papon, Guo}:
\begin{equation}
\chi = \frac{\alpha-\alpha^{\rm coil}}{\alpha^{\rm collagen}-\alpha^{\rm coil}}
\end{equation}

where $\alpha^{\rm coil}$ is the rotation angle measured in the coil state above $T_g$ and extrapolated down to $T$ according to \cite{Guo}, and $\alpha^{\rm collagen}$ is the rotation angle of a 1 cm-thick,  5 wt.\% collagen sample  \cite{Papon}. 

The sample cell is thermalized by a water circulating circuit.  Quenching is achieved by switching the circulating water between  two different temperature-controlled baths.  Although the total duration of the quench  is comparable to the one achieved in the rheometer, it is impossible to match precisely the shapes of the transients, the initial cooling rate being faster in the polarimeter. 
In the following we will chose the time origin for physical gelation as the instant at which the sample temperature reaches the target one $T$. 

\section{Results and discussion}

\begin{figure}[h]
\begin{center}
\includegraphics[width = 8 cm]{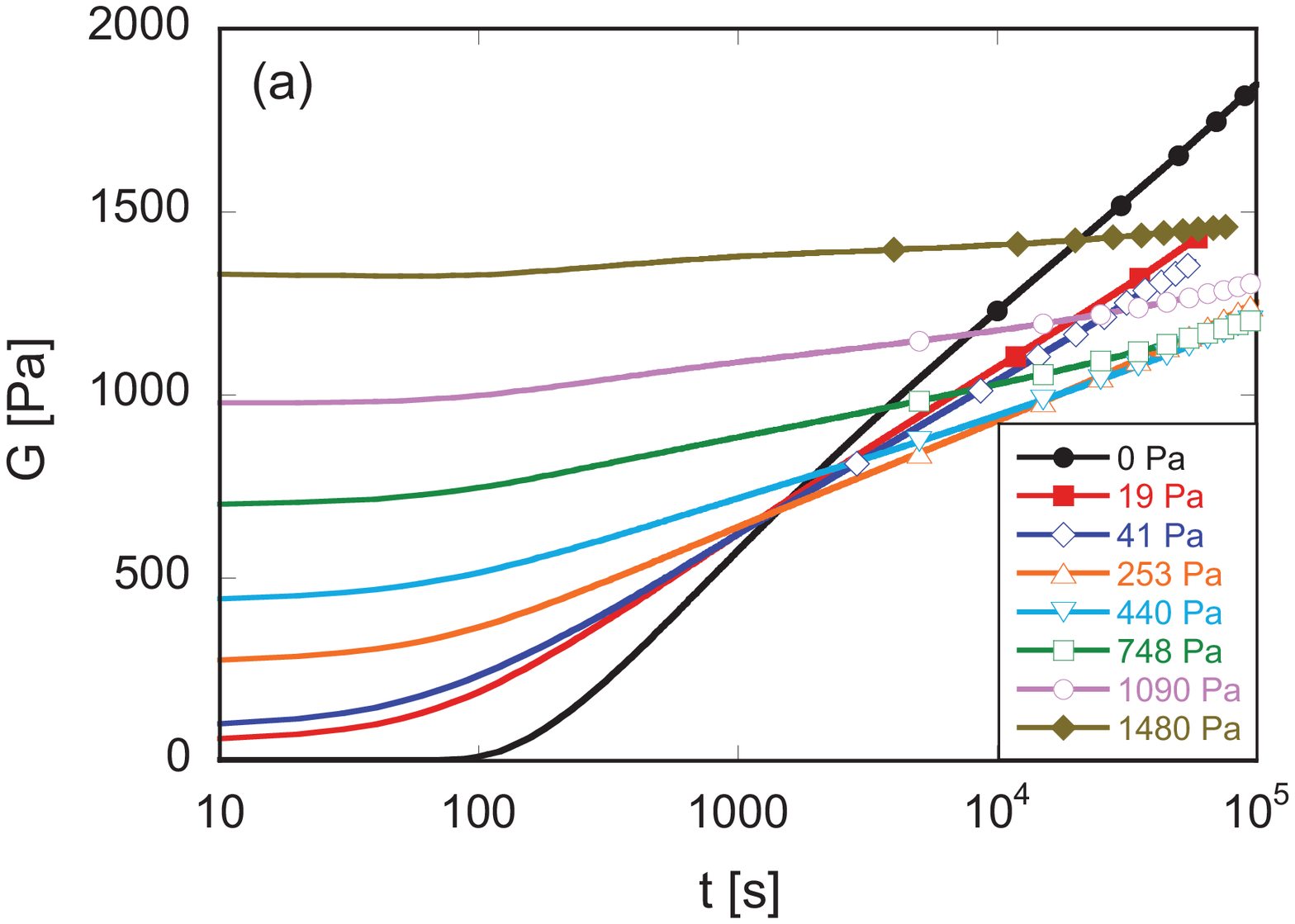}
\includegraphics[width = 8 cm]{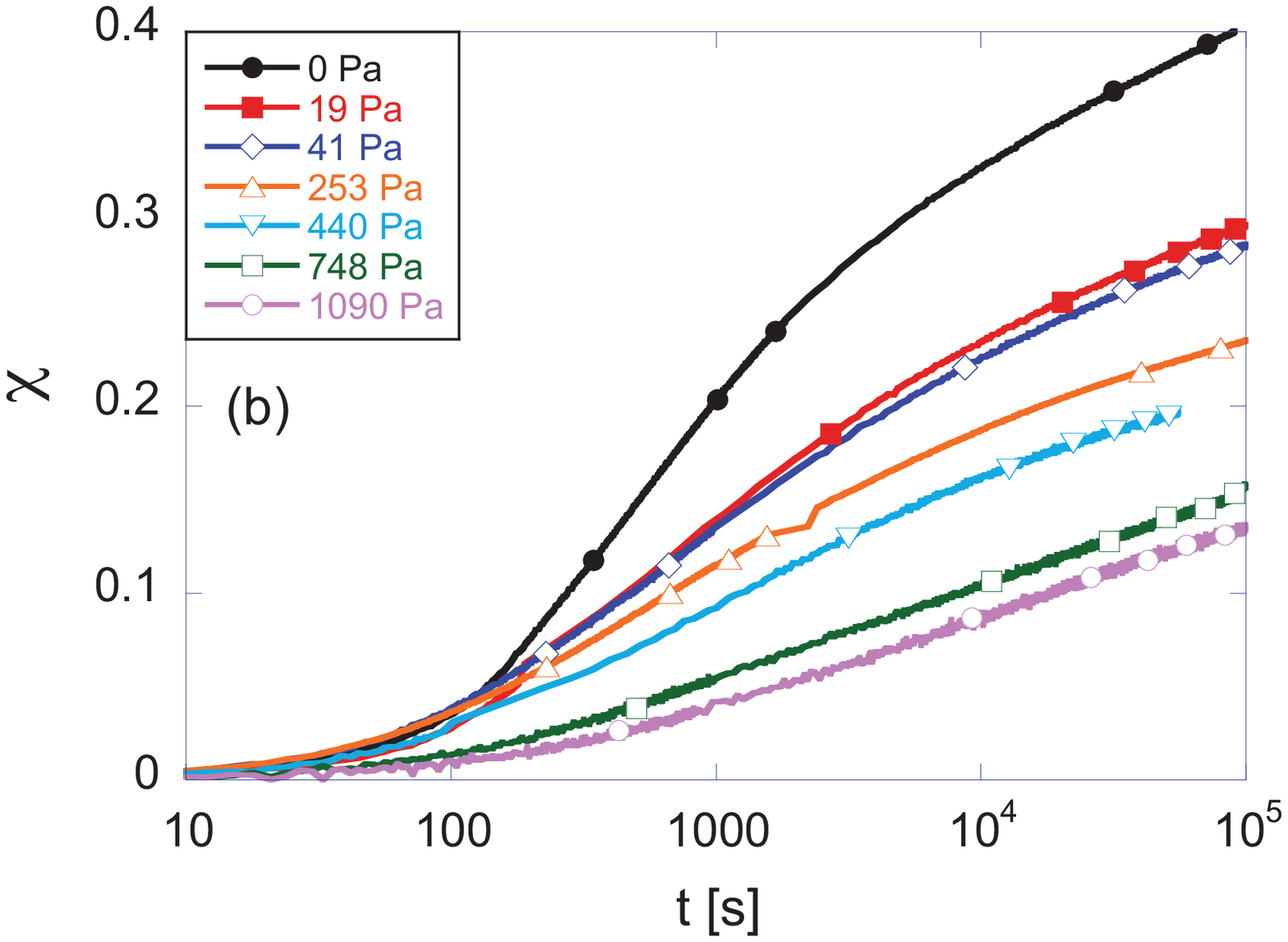}
\caption{(a) Shear modulus dependence on time after quench to $T=20^\circ$C of a set of hybrid gels with various densities of covalent crosslinks. The labels indicate the values of the shear modulus  $G_{ch}$ of the corresponding chemical networks.  
(b) Evolution of the renatured collagen fraction $\chi$ for the same gels. Data from the $G_{ch} = 1480$ Pa gel  have been discarded due to the smallness of the corresponding signal to noise ratio. }
\label{Fig:AllGchi}
\end{center}
\end{figure}

Figure \ref{Fig:AllGchi} shows the time evolution of the shear modulus $G$ and helix fraction $\chi$ for  values of the modulus of the covalent network at the working temperature $T = 20^\circ$C, increasing from $0$ (purely physical gel) to  $1480$ Pa.   
The semi-logarithmic representation (Fig.\ref{Fig:AllGchi}.a) reveals that the evolution of $G$ is qualitatively preserved in the presence of  covalent CL. In particular, late quasi-logarithmic aging is observed for all values of $G_{ch}$, no trend to saturation being observable up to $10^5$ s.

\begin{figure}[h]
\begin{center}
\includegraphics[width = 8 cm]{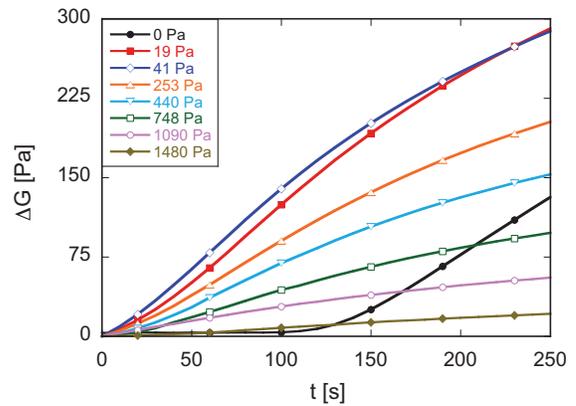}
\caption{Early evolution of the excess shear modulus $\Delta G(t) = G(t)-G_{ch}$. Same data as Fig.\ref{Fig:AllGchi}.a. }
\label{Fig:GZoom}
\end{center}
\end{figure}

The early  evolution of the modulus is best displayed by plotting the increment $\Delta G$ of its value from that immediately after quench.   The most conspicuous feature of the corresponding plot (Fig.\ref{Fig:GZoom}) consists of  the  initial behavior of $\Delta G$:
while the purely physical gel exhibits a finite delay before the emergence of a measurable modulus, for all hybrid gels, however small $G_{ch}>0$, the initial rate of growth   of $\Delta G$ is finite, decreasing as $G_{ch}$ increases. 
The origin of this discontinuous behavior is clear: while, in the purely physical gel case, percolation of  the triple-helix CL has to be reached for a finite $G$ to emerge, in the presence of a preexisting percolating covalent network, physical CL formation is immediately revealed. This leads us to an important inference, namely  a fraction at least of the triple-helix  CL involve polymer strands belonging to the covalent network. This proves that our hybrid gels  {\it do not} consist (as is the case of  ``double network'' ones \cite{GongDN}) of two interpenetrating but independent networks of different natures --- which would contribute additively to $G$. 
Finally,  the decrease of the initial slope indicates that covalent interchain bonds  do not act as inhomogeneous nucleation centers for triple helix formation. 

The evolution of the helix fraction $\chi$ is shown on Fig.\ref{Fig:AllGchi}.b. Here again the hindering effect of covalent CL translates into the gradual slowing-down of the dynamics with increasing CL density.  
Unlike $G$, the optical activity of the   purely physical gel  does not exhibit any latency since it is insensitive to the connectivity of the polymer chain system.  

Beyond these general remarks we need to define a procedure to compare modulus build-up in our various gels. We choose to proceed as follows: we first measure the time $t_0(G_{ch})$  for which the modulus of the purely physical gel $G_{ph}(t)$ reaches the value $G_{ch}$, then we shift the $G(t; G_{ch})$ curve of the corresponding hybrid gel by $t_0(G_{ch})$ along the time axis. 

The rationale for this choice is the following. A purely physical gel and a chemical one exhibit distinct network topologies, since  triple-helix CL are spatially extended, involving three chains,  while the pointlike  covalent ones connect two chains only. However,  as gel elasticity is of purely entropic origin, $G$ is controlled by the entropy reduction of the polymer chain system due to crosslinking, hence is an indicator of the density of degrees of freedom which are frozen at the corresponding stage of gelation \cite{MarquesJones}. So, our procedure amounts to choosing for the common reference the situation where   the  density of degrees of freedom which remain available for further evolution for both gels is comparable. 

\begin{figure}[htbp]
\begin{center}
\includegraphics[width = 8 cm]{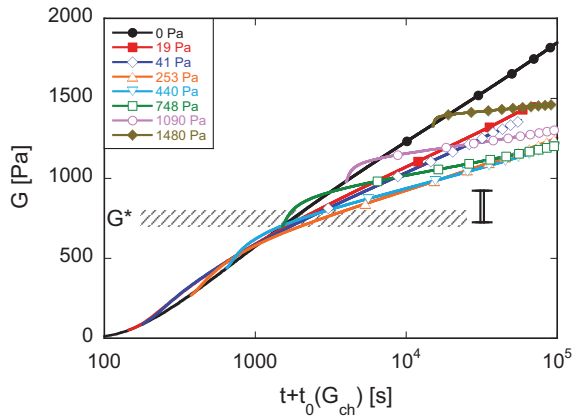}
\caption{ Same data as Fig.\ref{Fig:AllGchi}.a.: each hybrid gel curve has been shifted along the time axis so as to allow comparison with the purely physical gel one $G_{ph}(t)$ (black curve, full dots). $t_0(G_{ch})$ is the time at which $G_{ph} = G_{ch}$.  The hatched strip indicates the end of the early regime. The double vertical bar indicates the  cross-over range as obtained from $\chi(G)$ data (see text section III.A.)}
\label{Fig:ShiftG}
\end{center}
\end{figure} 

The resulting plot (Fig.\ref{Fig:ShiftG}) exhibits a striking structure. Namely, up to a modulus level $G^\star$ lying in the 700--800 Pa range,   the hybrid gel data corresponding to $G_{ch}\lesssim G^\star$ closely bunch with $G_{ph}(t)$.  Beyond this, they separate from $G_{ph}$ and bend sharply into a splayed set of quasi log-linear curves. For $G_{ch} \gtrsim G^\star$, separation occurs for $G-G_{ch}$ values which decreases rapidly with $G_{ch}$.

This structure leads us to identify,  in the network build-up process, two clearly different regimes  separated by a cross-over:

(i) an early stage  in which the dynamics is basically insensitive to the nature of preexisting CL. 

(ii) a late, aging one in which the modulus growth is strongly affected by the degree of CL reversibility as revealed by the marked decrease  with increasing $G_{ch}$ of the logarithmic slope. 

We now discuss separately in more detail  the features of these two regimes. 

 \subsection{Early stage} 
 
The bunching of the $G$ curves (Fig.\ref{Fig:ShiftG}) suggests that,  in this regime, it is the total gel modulus which  codes the network evolution. If such is the case, other characteristics of the network should exhibit a change of regime at a the crossover value comparable to $G^\star$. A natural candidate for checking this proposition is the helix fraction $\chi$ which we therefore plot, in the ``master curve''  representation of Joly-Duhamel et al.  \cite{Master}, as a function of $G$.

 \begin{figure}[htbp]
\begin{center}
\includegraphics[width = 8 cm]{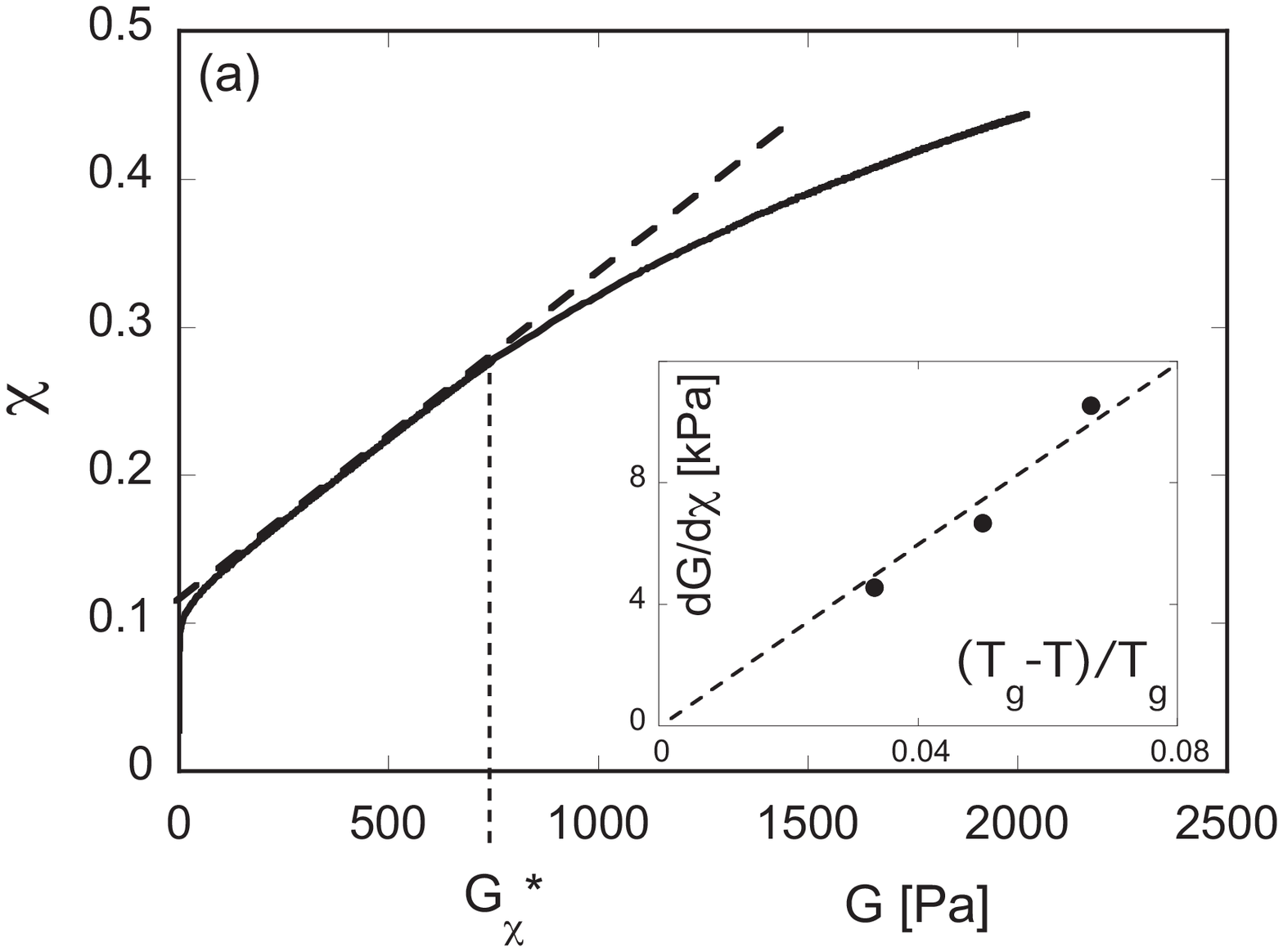}
\includegraphics[width = 8 cm]{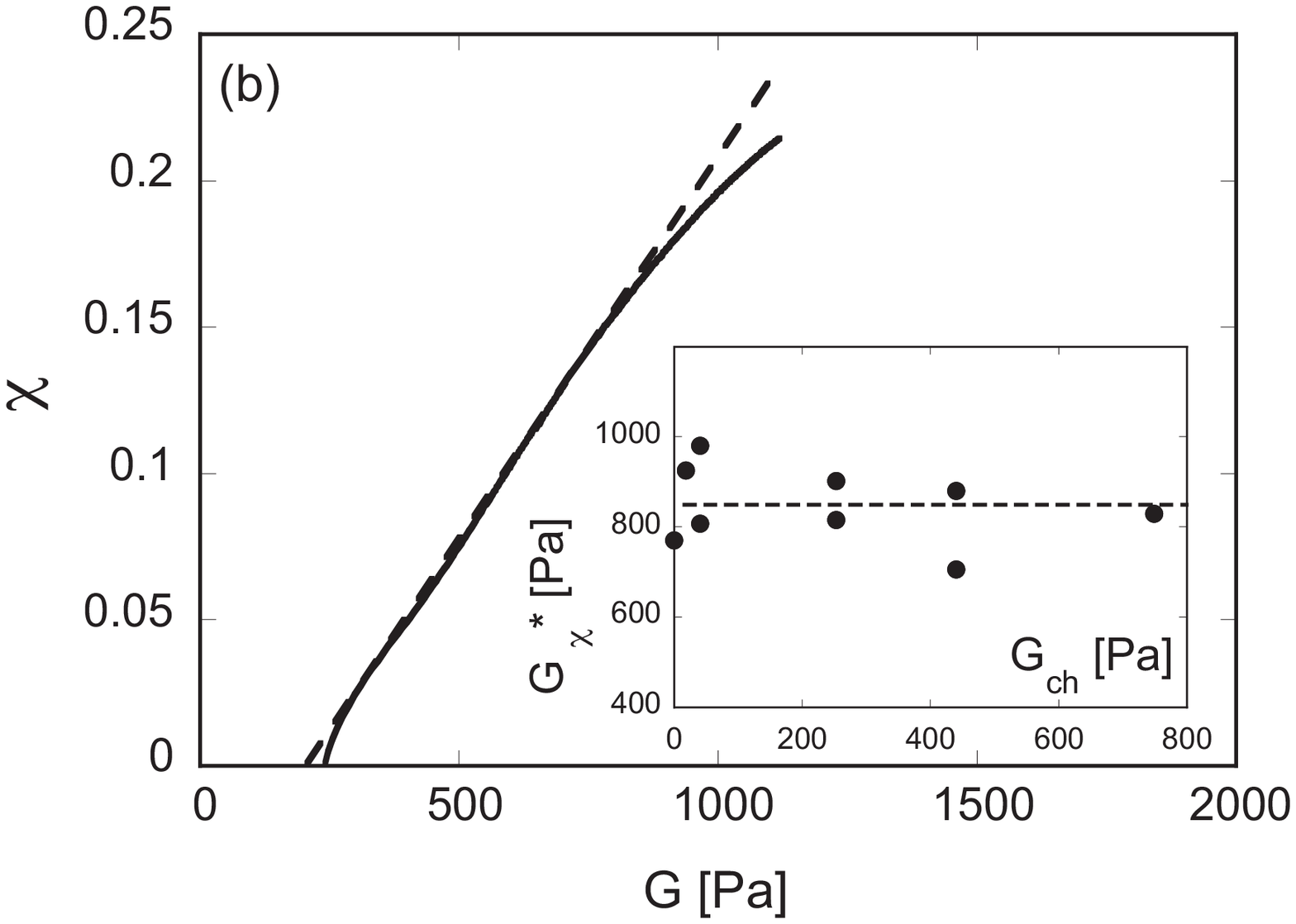}
\caption{(a) Helix fraction $\chi$ vs. shear modulus $G$ for the physical gel at $T = 20^\circ$C. The thick dashed line is the fit to the initial  linear regime ending at $G = G_\chi^\star$. Insert: Inverse of  the linear regime slope,   $dG/d\chi$, vs. reduced departure from gel temperature $T_g$. 
(b) Helix fraction versus shear modulus for a hybrid gel with $G_{ch} = 253$ Pa at $T = 20^\circ$C. Insert: End of linear regime $G_\chi^\star$ vs. chemical modulus $G_{ch}$ at the same temperature.  }
\label{Fig:ChiG}
\end{center}
\end{figure}

Fig.\ref{Fig:ChiG}.a shows the corresponding plot for a purely physical gel at $T = 20^\circ$C. The finite $\chi(G=0)$ value measures the helix fraction at percolation. At the modulus value noted $G^\star_\chi = 770\pm 40$ Pa, $\chi(G)$ crosses over from a linear regime to a markedly sublinear one. The same analysis of the hybrid gel data for $G_{ch}$ values up to 750 Pa reveals, as illustrated on Fig.\ref{Fig:ChiG}.b, the persistence of a low-$G$ linear regime terminating at $G^\star_\chi(G_{ch})$. We find (see insert of Fig.\ref{Fig:ChiG}.b) that $G^\star_\chi$ does not exhibit any significant dependence on $G_{ch}$. The mean ($\pm$ std. dev.): $\bar G_\chi^\star = 830\, (\pm 90)$ Pa is fully compatible with the crossover level $G^\star$ previously identified from the bunching of the $G(t)$ data (see Fig.\ref{Fig:ShiftG}). 
Moreover, the $G^\star_\chi$ crossover turns out (Fig.\ref{Fig:loglin}) to lie very close to the emergence of the logarithmic  aging regime. In order to check the robustness of these features with   respect to temperature we have performed the same analysis on physical gels of the same composition, quenched at $T = 15$ and $10^\circ$C. As is well known, the smaller $T$, the faster the gelation dynamics. The shape of the $\chi(G)$ curves  is preserved while $\bar G^\star_\chi$ increases as $T$ decreases --- its values, reported in Table \ref{Tab:Moduli} ,  remaining in the close vicinity of the entry into the log stage (see Fig.\ref{Fig:TroisTemp}). 

\begin{table}[htdp]
\caption{default}
\begin{center}
\begin{tabular}{|c|c|c|}
\hline
$T\,[^\circ$C]&$\bar G_\chi^\star$ [Pa]&$\mathcal G$ [Pa]\\
\hline
10&2650&2016$\pm$310\\
15&1670&1300$\pm$160\\
20&830&685$\pm$15\\
\hline
\end{tabular}
\end{center}
\label{Tab:Moduli}
\end{table}

These results bring strong support to our above assumption of the existence of an early regime, the development and termination  of which are  coded by the total elastic modulus, i.e. are quasi-insensitive to network topology and crosslink  reversibility.  

Moreover, the quasi-linearity of the early $\chi(G)$ dependence suggests a simple, qualitative interpretation, namely: according to the entropic origin of gel elasticity, it is commonly admitted  \cite{Rubinstein} that, as a first approximation,  the shear modulus  scales as the CL density. During physical gelation, only triple helix CL are formed. So, on the same level of approximation, we may write for the variations of $G$ and $\chi$ induced by an increment $\Delta\nu$ of CL density: $\Delta G\sim k_BT\Delta\nu$ and $\Delta\chi\sim \bar\ell\Delta\nu$ with $\bar\ell$ the average triple-helix CL length. From this,    $d\chi/dG=$ cst  immediately translates into the statement that $\bar\ell$ remains stationary over the early gelation regime. 
Note that this assertion agrees with the conclusion reached by Guo et al \cite{Guo} who were able to deduce, from the evolution of $\chi$ upon remelting physical gels of various ages, the distribution of lengths of these rods at various stages of  gelation. They found that it remains quasi invariant in the corresponding time regime (while $\bar\ell$ slowly drifts, at later times, toward larger values). 
 
 One step further, in agreement with nucleation theory \cite{FloryWeaver}, we can expect the critical nucleus size, hence $\bar\ell$ to diverge upon approaching the gelation temperature $T_g$. Since, from our above remarks, $1/\bar\ell \sim dG/d\chi$, we have plotted $dG/d\chi$ vs. $(T_g-T)/T_g$.  We find that our data for $T = 10, 15, 20^\circ$C extrapolate  linearly to the origin (see insert of Fig.\ref{Fig:ChiG}.a), indicating  the expected $(T_g-T)^{-1}$ divergence. 
 
\subsection{Late stage}

\begin{figure}[htbp]
\begin{center}
\includegraphics[width = 8 cm]{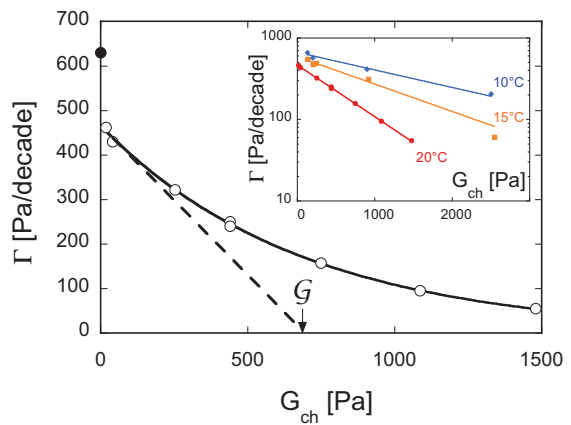}
\caption{Logathmic slope $\Gamma  = dG/d\log_{10} t$ vs. $G_{ch}$ for hybrid gels at $T = 20^\circ$C. Full line: best fit   $\Gamma = A\exp(-G_{ch}/\mathcal G)$ with $\mathcal G = 685$ Pa. The purely physical gel data point (full dot) has been excluded from the fit (see text).  Insert: $\Gamma$ vs. $G_{ch}$ at different temperatures below $T_g$. }
\label{Fig:Pentes}
\end{center}
\end{figure}

This regime is characterized by the logarithmic increase with time of the shear modulus which emerges close above the crossover level $G^\star$ defined on Fig.\ref{Fig:ShiftG}. As already mentioned, the logarithmic slope:
\begin{equation}
\label{Eq:Gamma}
\Gamma =  \frac{d G}{d(\log_{10} t)}
\end{equation}
is markedly dependent upon $G_{ch}$, i.e. on the density of irreversible CL.  Fig.\ref{Fig:Pentes} displays the values of $\Gamma$ measured at $T=20^\circ$C for various  hybrid gels with $G_{ch}$ ranging from 20 to 1600 Pa. These data are very nicely fitted by the expression
\begin{equation}
\label{Eq:SlopeFit}
\Gamma = \Gamma_0\exp\left[-\frac{G_{ch}}{\mathcal G}\right]
\end{equation}
with  $\mathcal G = 730\pm15$ Pa.
Note that the extrapolated slope $\Gamma_0 = 466\pm4$ Pa/decade lies noticeably below the value (630 Pa, see black dot on Fig.\ref{Fig:Pentes}) for the purely physical gel.  This is attributable to the fact that, while the physical gel does not contain any irreversible CL, a hybrid gel with vanishingly small $G_{ch}$ presents the finite density of covalent CL corresponding to the percolation threshold of the chemical network.
The above exponential decay is robust with respect to temperature variations as shown on the insert of Fig.\ref{Fig:Pentes}, which reports data obtained at $T = 10$ and $15^\circ$C.

\begin{figure}[htbp]
\begin{center}
\includegraphics[width = 8 cm]{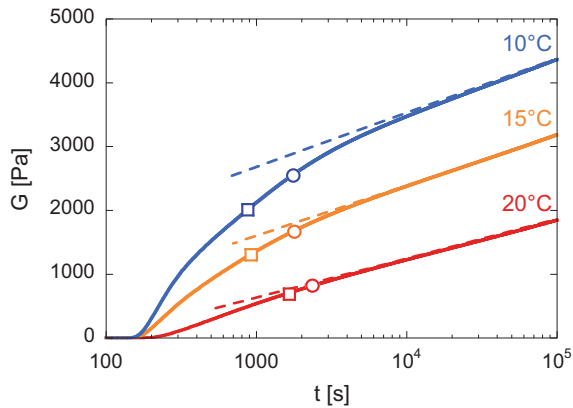}
\caption{Modulus build-up curves $G_{ph}(t)$ at different temperatures. The circles indicate the cross-over modulus $\bar G_\chi^\star$. The squares indicate the  modulus $\mathcal G$ characteristic of the sensitivity of gel aging to $G_{ch}$ (see Table \ref{Tab:Moduli})}
 
\label{Fig:TroisTemp}
\end{center}
\end{figure}

This behavior brings to light the existence of a characteristic modulus $\mathcal G(T)$ associated with logarithmic aging. We notice that its variation with $T$  parallels that of the crossover modulus while its values remain  systematically smaller than $G_\chi^\star$  (see Table \ref{Tab:Moduli} and Fig. \ref{Fig:TroisTemp}).

\subsection{Discussion}

These results demonstrate the existence of two dynamical regimes during the  build-up of the physical network, which can be clearly differentiated according to their sensitivity to the nature of crosslinks. 

While covalent CL are irreversible, physical ones  undergo ``rearrangements'' via elementary, thermally activated zipping or unzipping events through which helix segments may  either grow, shrink or slide by coherent zipping/unzipping at both ends. The clock is set  for these fluctuations by the cis-trans isomerisation of (Pro-Gly-Pro) sequences of amino acid residues \cite{Nij, McBride, Bachinger}. As the corresponding  mean flip time lies in the 10--100 s range at  $T\simeq T_g$, we are dealing here with very slow processes. Rearrangements occur via the biasing of fluctuations under the effect of  tension forces along the coil segments which connect the CL. 

In the early regime, the evolutions of  both the  stiffness and the  helix fraction  are found to be coded by the instantaneous value of the modulus and  essentially insensitive to  the presence of covalent bonds. This implies that the time scale for the nucleation of triple helices  is small enough for rearrangements to be quasi-frozen, hence the insensitivity to the (reversible or not) nature of the CL. 

Of course, the larger $G$, the shorter the mesh size hence the thinner the entropic supply available for further nucleation the rate of which thus decreases --- so much so that rearrangement dynamics finally takes over.  
This results in a  rather sharp switch, about a crossover modulus $G^\star$. The emerging  late regime is  characterized by the   logarithmic growth of $G(t)$ together with a markedly sublinear $\chi(G)$ --- which indicates that   less and less new helix length is required for further stiffening of the network. Moreover, and most important, the aging dynamics  dramatically slows  down as the density of irreversible CL increases. All these features are consistent with a dynamics controlled by CL rearrangements.

Loosely speaking we may distinguish between two types of such processes, which we refer to as {\it local} and {\it collective}. We define as a local process the length evolution of a triple helix segment in the frozen environment determined by its non-evolving nearest neighboring CL. Due to the finite, nanometric, size of the gel mesh,  the relaxation time for such a process is clearly finite. A simple model of the helix-coil transition on a single polymer strand with fixed ends \cite{Kutter} and with a Fokker-Planck dynamics  \cite{PGG} for the  helix length evolution yields a linear dependence of the relaxation time on the number of monomers in the strand \cite{Caroli}. Explaining the late dynamics by the ``local'' growth of triple helices nucleated in the early stage would demand a flat distribution of relaxation times spanning more than the 3 decades over which logarithmic aging has been observed, hence a completely unrealistic   flat  (gaussian) strand end-to-end distance  distribution  spanning at least 1.5 decades. 

We are therefore led to conclude that the aging behavior results from  dynamical interactions between  crosslinks, which we call collective processes. More precisely: an elementary (un)zipping event affecting a physical CL  induces variations of the tensions of emerging strands, i.e. acts as a localized  force dipole acting on the surrounding elastic  medium. This gives rise quasi-instantaneously (i.e. with a delay set by sound propagation)  to a power law decreasing strain field which, in turn, modifies the tensions of strands involved in remote CLs, thereby displacing their equilibrium triple helix length. Due to the long range of elastic couplings, this results in a broadband  mechanical noise.
  As global relaxation, i.e. collagen renaturation, proceeds,  one expects that further modulus growth requires cooperation between an increasingly large number of physical CL, hence involves a growing correlated volume and, accordingly, a growing time scale.    
  
  In such a picture, introducing an increasing  fraction of irreversible CL has a  multiple effect. On the one hand, for a given $G$ it reduces the fraction of rearrangeable CLs. On the other hand, the average distance between the latter increases, and so does the mechanical noise amplitude.  Moreover, covalent bonds restrict the amplitude of triple helix sliding due to coherent zipping/unzipping at both ends. We saw that as time lapses in the late regime, modulus increase involves  less and less helix length creation. This is a strong hint of the  increasing role played by  length preserving correlated sliding processes, which are destroyed by  the interposition of covalent bonds on chains connecting the two helix segments. This topological effect of chemical bonds thus results in the decimation of the possible relaxation paths in configuration space.

 At this stage, pending the development of a theory relating  logarithmic aging to definite characteristics of structural evolution, it  remains out of the scope of the present work to  explain on the basis of  these qualitative remarks  the observed dependence of the aging dynamics of hybrid gels on their chemical modulus $G_{ch}$. Progress toward this goal would certainly highly benefit from the development of numerical modeling, possibly in the spirit of the investigations of long time scale glass aging currently developed by several groups \cite{Mousseau, Barkema, Egami}. 
 
This might shed light on an issue of broader interest, namely whether or not stiffening during aging is associated with decreasing elastic non-affinity, i.e. whether slow relaxation drives the system toward gradual homogenization of the internal stresses
 
\subsection*{Acknowledgement}   We acknowledge funding from Emergence-UPMC-2010 research program.

\end{document}